\documentclass[3p]{elsarticle}
\usepackage{hhline}
\usepackage{fancyvrb}
\newcounter{bla}
\usepackage[usenames,dvipsnames]{color}

\journal{Computer Physics Communications}

\begin{document}

\begin{frontmatter}
\title{Variable-step-length algorithms for a random walk: hitting probability and computation performance}
\author[SCC,HSE]{Olga Klimenkova}
\author[SCC]{Anton Yu. Menshutin}
\author[HSE,Landau]{Lev N. Shchur}
\address[SCC]{Science Center in Chernogolovka,142432 Chernogolovka, Russia}
\address[HSE]{National Research University Higher School of Economics, 101000 Moscow, Russia}
\address[Landau]{Landau Institute for Theoretical Physics, 142432 Chernogolovka, Russia}

\begin{abstract}
We present a comparative study of several algorithms for an in-plane random
walk with a variable step. The goal is to check the efficiency of the
algorithm in case where the random walk terminates at some boundary. We
recently found that a finite step of the random walk produces a bias in the
hitting probability and this bias vanishes in the limit of an infinitesimal
step. Therefore, it is important to know how a change in the step size of
the random walk influences the performance of simulations. We propose an
algorithm with the most effective procedure for the step-length-change
protocol.
\end{abstract}
\end{frontmatter}

\section{Introduction}

Simulation of a random walk is a very general approach in many areas of
science and engineering, for example, in physics (cover of a
torus~\cite{Grassberger2017}), biology (leukocyte
migration~\cite{Jones2015}), chemistry (formation of crystal
patterns~\cite{Srivastava2016}), health research (human
growth~\cite{Suki2017}), earth science (selection of river
networks~\cite{Rinaldo2013}), and natural resources research
(geostatistics~\cite{Caixeta2017}) to mention just a few research areas.

A random walk in a domain is simulated with a finite step, i.e., with jumps
of the walker of some finite distance. The size of the jumps is irrelevant
while the walker is far from the domain boundary, and there is a well-established method to speed up simulations
using the large step size far away from the domain~\cite{Ball85}. The efficient algorithm to control distance to the domain boundary
is based on the marked hierarchical memory (see algorithm~\cite{Tol-Meak} for the lattice walk  and algorithm~\cite{MSV} for the off-lattice walk),
and a proper procedure for changing the size of the jumps when close to the domain
boundary must be chosen. Realization of such algorithm for the contemporary computers with relatively big onboard memory published in ~\cite{KMO}.
Anyway, the last jump to the boundary domain is always finite in all known methods and algorithm realizations. 

It was recently found~\cite{KLM-1} that the finiteness of the size of random walk jumps produces a visible bias in the
hitting probability. The walker moving in the plane from infinity hits the
circle at the origin and the bias in the hitting probability depends on the
angle between the position of the hitting point and the radius at which the
walker starts.\footnote{The problem of estimating the accuracy of the probability of the error in Monte Carlo simulations was emphasized in the very early paper of Metropolis and Ulam on the subject entitled ``The Monte Carlo Method'' (see the last two sentences of the next-to-last paragraph in the paper~\cite{TheMCmethod}).} Fortunately, the bias vanishes in the limit of an infinitesimally small step size. This motivates the present study of the
efficiency of simulations while varying the step size using different protocols.
Simulating a random walk with a very small size is impractical, and some protocol for changing the
size must be implemented. 

In this paper, we check how different protocols can influence the
simulation efficiency, minimizing the time needed to hit the boundary. 
We estimate numerically using different protocols the probability  for in-plane random walk to hit the circle placed at origin.
The probability is known exactly, and it was found in the paper~\cite{KLM-1}  that the bias have maximum absolute value at zero angle (and at the angles $\pm\pi)$ with respect to the initial position of the random walk, and that the bias vanishes with vanishing  jump size. In the present paper we choose the bias at zero angle as indicator of the accuracy of the estimated hitting probability.

The paper is organized as follows. In section~\ref{sec-Model}, we introduce the
model of the random walk in the plane and provide exact results for the
termination probability. In section~\ref{sec-protocols}, we discuss the
basic algorithm, introduce the observables to control accuracy for the
hitting probability, and propose the three different protocols for the
variable step of the random walk. In section~\ref{sec-simulations}, we
present the results of the simulations. A short discussion of the results
in section~\ref{sec-discussion} concludes our paper.

\section{Model}
\label{sec-Model}

One of the most interesting cases for simulating a random walk is the
random walk in the plane, for at least two reasons. First, it is well
defined in the sense that the probability to escape to infinity is zero. The
unbiased random walk is fully ergodic, it visits an $\epsilon$-neighborhood
of any point in the limit of infinite time. The technical problem is that
the time to reach such a neighborhood is logarithmically divergent with
$\epsilon{\rightarrow}0$. Fortunately, the problem of infinite time can be
eliminated because of the second reason, the existence of an exact formula
for the hitting probability. The probability is defined in terms of the
kernel solution of the corresponding two-dimensional Laplace
equation~\cite{Free-SSZ,KVMW,Killing-free}.

\begin{figure}[ht]
\begin{center}
\includegraphics[scale=0.5]{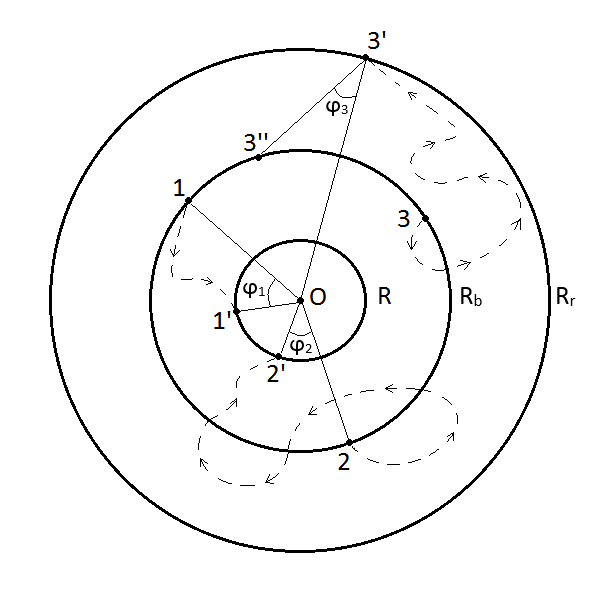}
\caption{The random walk starts at the birth circle of radius $R_b$. The
particle terminates when it hits the circle of radius $R$. If the particle
crosses the returning circle $R_r$, then it is placed on the birth circle
at the corresponding angle (see expression~(\ref{eq-return}) and the
discussion in the text).}
\label{fig:walk-scheme}
\end{center}
\end{figure}

We perform simulations in the following geometry. An absorbing circle of
radius $R$ is placed at the origin, and the walker starts at any point on a
``birth'' circle of radius $R_b$. Walkers terminate at the absorbing circle
$R$. Figure~\ref{fig:walk-scheme} schematically shows two trajectories of
this type, trajectory 1--$1'$ and trajectory 2--$2'$.

The exact probability to hit a point on the absorbing circle $R$ is given by
\begin{equation}
P(\phi) = \frac{1}{2\pi} \frac{x^2-1}{x^2-2x\cos \phi +1},
\label{eq-return}
\end{equation}
where $x=R_b/R>1$ and the angle $\phi$ is measured between the radius of the
initial position of the walker on the ``birth'' circle $R_b$ and the radius
of the hitting point on the absorbing circle $R$. The angles $\phi_1$ and
$\phi_2$ for trajectories 1--$1'$ and 2--$2'$ are shown in
Fig.~\ref{fig:walk-scheme}.

To reduce the computational time, we prevent walker from going far away: if
it goes farther than the distance $R_r$ from the origin, then we return the
walker to the birth radius at the angle $\phi$ calculated using the
probability given by expression~(\ref{eq-return}) with $x=R_r/R_b>1$
(see~\cite{Killing-free} for details). This case is plotted in
Fig.~\ref{fig:walk-scheme} as the trajectory 3--$3'$, which generates a
walker at the point $3''$ at the radius $R_b$ with the angle $\phi_3$. The
angle $\phi$ with distribution~(\ref{eq-return}) is generated using the
expression~\cite{Free-SSZ, Sander2000,Killing-free}
\begin{equation}
\phi = 2\arctan \left( \frac{x-1}{x+1} \tan u\frac{\pi}{2} \right)
\label{eq-return}
\end{equation}
with a random variable $u$ uniformly distributed in the interval $[-1,1]$.

We must stress that this is not only a computational trick but also the way
to include the infinite boundary condition {\em exactly} for the solution of
the Laplace problem in the plane. Using this ``killing-free'' algorithm in
diffusion-limited aggregation simulations, we never observe instability of
the DLA cluster, which is the case in simulations in which the walker is
simply removed after crossing the circle of radius $R_r$. The finite ratio
of $R_r$ to $R_b$ leads to a distortion of the infinite boundary conditions
and generates a Saffman--Taylor instability~\cite{ST-1958,LTMS-2010}, due to
which the DLA cluster grows in only one direction~\cite{Killing-free} and
develops only one of the branches.

\section{Algorithm and protocols}
\label{sec-protocols}

First, we check how the accuracy of estimating the hitting probability and
the computation time depends on the jump size of the random walk. We perform
$N$ random walks, typically $N=10^5$ to $10^6$. For each of $N$ random walks,
we generate a random angle $\phi_0$ uniformly distributed in $[-\pi:\pi]$ as
the initial coordinate on the circle of radius $R_b$ (we define the
direction of the angles clockwise and the value of the angles from the
horizontal line). At each jump of the walk, we generate a random angle
$\psi$ associated with the direction of the jump at the distance $\delta$.
We calculate an estimate of the hitting probability $P_{\mathrm{sim}}(\phi)$
by dividing the interval $[-\pi:\pi]$ of possible hitting angle values
$\phi$ into 180 bins and counting the number of hits for each bin.
Normalizing the results over the total number $N$ of random walkers and over
the bin size gives the estimate of the hitting probability
$P_{\mathrm{sim}}(\phi)$. The deviation of the estimate from the exact
result is calculated as
\begin{equation}
\Delta P(\phi)=\frac{P_{\mathrm{sim}}(\phi)-P(\phi)}{P(\phi)}.
\label{sim-prob}
\end{equation}

Repeatedly estimating $\Delta P_i(\phi)$ with $N$ walkers $M$ times provides
the average deviation $\Delta P_{\mathrm{av}}(\phi)$ and its standard error $D(\phi)$:
\begin{eqnarray}
&&\Delta P_{\mathrm{av}}(\phi)=\frac{\sum_{i=1}^M \Delta P_i(\phi)}{M},
\label{sim-D}
\\
&&D(\phi)=\frac{\sqrt{\sum_{i=1}^M \left(\Delta P_i(\phi)-\Delta P_{\mathrm{av}}(\phi)\right)^2}}{M}.
\label{sim}
\label{error-d}
\end{eqnarray}

\begin{figure}[!ht]
\begin{center}
\includegraphics[scale=0.5]{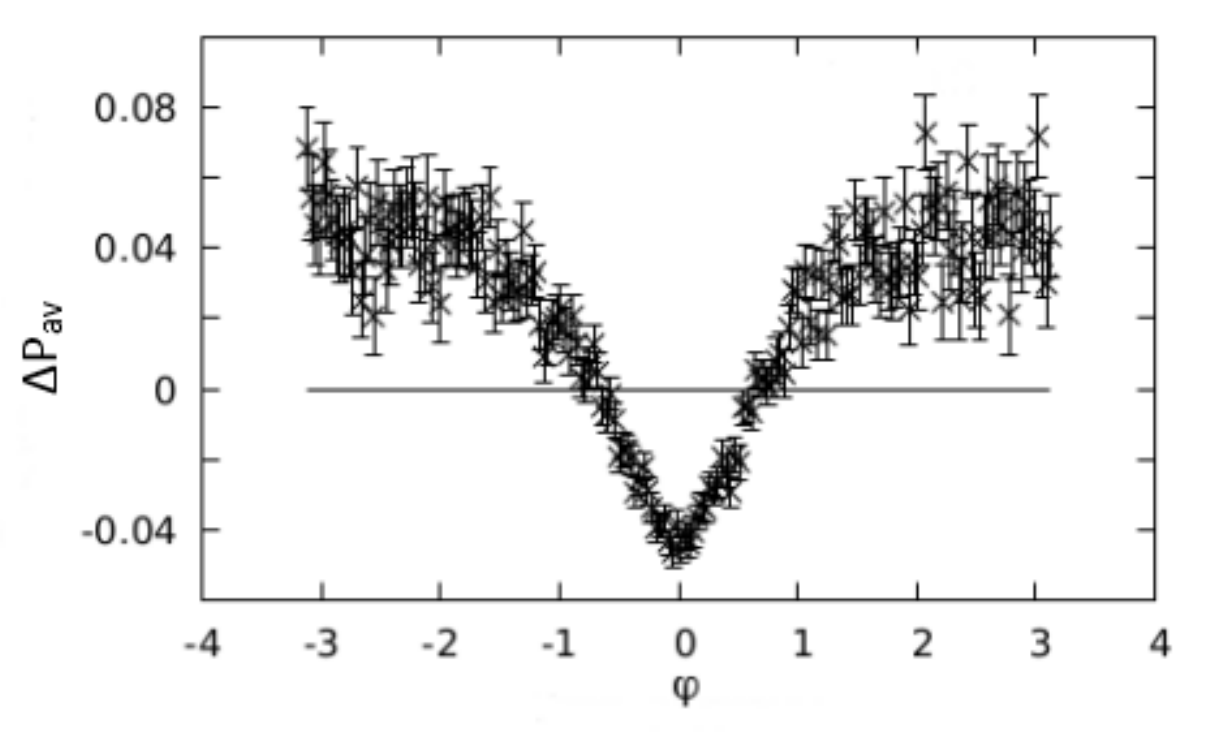}
\caption{Deviation $\Delta P_{\mathrm{av}}(\phi)$ for the random walk step
length $\delta=1$ with $M=100$.}
\label{fig:deviation}
\end{center}
\end{figure}

It was observed in our previous paper~\cite{KLM-1} that i) the deviation depends on the angle $\phi$ as shown in Fig.~\ref{fig:deviation} and ii) the deviation has an extremum at $\phi=0,\pm\pi$.  It was proposed~\cite{KLM-1} that estimated probability does depend on the angle $\phi$ and jump size $\delta$ as 

\begin{equation}
P_{\mathrm{sim}}(\phi)\approx P(\phi) \left(1-\left(\frac{\delta}{R}\right)^{\alpha} cos\;{\phi}\right)
\label{expression}
\end{equation}
with positive exponent $\alpha<1$.
 
It is visible from the Figure~\ref{fig:deviation} and from the Expr.~\ref{expression} that the maximum deviations occur at the angles $\phi=\pm \pi$ and 0.
We choose the value $\Delta P_{\mathrm{av}}=\left|\Delta P_{\mathrm{av}}(0)\right|$ as an indicator of the accuracy of the estimated hitting probability $P_{\mathrm{sim}}(\phi)$ because it has lower dispersion in comparison with the values at the angles $\pm\pi$.

It was shown in~\cite{Killing-free} that finite-size effects are not visible
for $R_r\gg R_b>R$. We therefore choose $R=10$, $R_b=20$, and $R_r=200$ in
the simulations. The hitting angle was calculated at the point of
intersection of the trajectory with the hitting circle $R$. For the value
$R=10$ used in the simulations, for jump values limited to $\delta\le1$, and
for the bin width $2\pi/180$, this choice of the angle does not produce a
visible bias.

We check three protocols.
\begin{itemize}
\item[P1:]{\em Equal jumps protocol.} Simulations are performed with a
fixed jump distance $\delta$.
\item[P2:]{\em Two regions protocol.} The inner space between the circles
$R$ and $R_b$ is divided into two regions. Jumps are performed with the
distance $\delta_1$ for a particle with the coordinate $r$ in the range
$R_c<r<R_r$ and with the distance $\delta_2$ for a particle with $r$ in
the range $R<r<R_c$. Accordingly, $\delta_2$ is taken smaller than
$\delta_1$.
\item[P3:]{\em Linear protocol.} The jump size $\delta$ of the random walk
is changed as a linear function of the distance $r$ to the origin:
\begin{equation}
\delta(r) =\delta_b-(\delta_b-\delta_h)\frac{R_b-r}{R_b-R}.
\label{eq:delta-linear}
\end{equation}
The jump value $\delta(r)$ hence decreases from $\delta_b$ at the birth
circle $R_b$ to the $\delta_h$ at the hitting circle $R$ and is larger
than $\delta_b$ when the walker travels outside the birth circle $R_b$.
\end{itemize}

\section{Simulations}
\label{sec-simulations}

\subsection{Equal jumps protocol P1}

The simulation results using protocol P1 are presented in
Table~\ref{tab-P1}, where the number in parenthesis shows the statistical
uncertainty to the last digits of the quantity. The influence of the step
length on the accuracy of the hitting probability is nicely demonstrated.
Indeed, the data in the table shows that simulations with the smaller
length size lead to better precision: a length ten times smaller gives a
three times better precision. At the same time, the computation time
increases by two orders of magnitude. The average number of jumps $K$ also
grows drastically as the step length decreases.

\begin{table}[!ht]
\centering
\caption{Computational performance for the equal jumps protocol P1.}
\label{tab-P1}
\begin{tabular}{|r|r|r|l|} \hline
$\delta$ & $\Delta P_{\mathrm{av}} $ & Time, min & $K$\\ \hline
\hline
1 & 0.056(2) & 0.1520(4) &4814(28) \\ \hline
0.1 & 0.015(2) & 14.68(3) & 445359(2860) \\ \hline
\end{tabular}
\end{table}

It can be seen from Table~\ref{tab-P1} that longer jumps give better
performance but a higher deviation of the hitting probability while shorter
jumps give a better quality of the hitting probability and longer
computation times. We must therefore find an optimum in the space of
performance--accuracy. In practice, we should choose a protocol of jump
length variation.

The optimal simulation should use longer jumps far from the absorbing domain
and shorter jumps close to it. We consider two implementations of this idea
in protocols P2 and P3.

\subsection{Simple protocol P2}

Protocol P2 is designed to check the idea that only final jumps influence
the precision of the hitting probability. In Table~\ref{tab-simple}, we show
a summary of the simulations for different values of $R_c$, $\delta_1$, and
$\delta_2$.

\begin{table}[!ht]
\centering
\caption{Performance and precision evaluation for the simple protocol P2.}
\label{tab-simple}
\begin{tabular}{|l|r|r|r|r|l|} \hline
$R_c$ & $\delta_1$ & $\delta_2$ & $\Delta P_{\mathrm{av}} $ & $T$, min & $K$ \\ \hline\hline
11 & 1 & 0.3 & 0.024(2) & 0.1559(2)& 4543(29) ) \\ \hline
& 1 & 0.1 & 0.021(2) & 0.1544(2) & 4555(27) \\ \hline
& 1 & 0.05 & 0.015(2) & 0.1529(3)& 4737(27) \\\hline
& 1 & 0.03 & 0.015(2) & 0.1629(2)& 5246(25) \\\hline
& 1 & 0.02 & 0.011(2) & 0.1695(2)& 6287(28) \\\hline
& 1 & 0.015 & 0.015(2) & 0.1933(4)& 7637(27) \\\hline
& 1 & 0.01 & 0.013(2) & 0.2507(2) & 11582(31) \\\hline
\hline
15 & 1 & 0.3 & 0.027(2) & 0.1624(2)	&	4983(26) \\\hline
& 1 & 0.1 & 0.020(2) & 0.2236(2)	&8351(30) \\\hline
& 1 & 0.05 & 0.013(2) & 0.4106(6)	&19878(43) \\\hline
& 1 & 0.03 & 0.012(2) &	0.8677(5)	&47027(77) \\\hline
& 1 & 0.02 & 0.016(2) & 1.754(1)	&100104(173) \\\hline
& 1 & 0.015 & 0.014(2) & 2.874(4)	&174131(312) \\\hline
& 1 & 0.01 & 0.014(2) &	 7.38(1)	&386544(635) \\\hline
\hline
15 & 5 & 0.3 & 0.027(2) & 0.00976(7)	&		379(1) \\\hline
& 5 & 0.1 & 0.017(2) & 0.03362(7)	&		1773(4) \\\hline
& 5 & 0.05 & 0.015(2) & 0.1116(1)	&		6402(13) \\\hline
& 5 & 0.03 & 0.014(2) &0.2962(2)	&		17289(35) \\\hline
& 5 & 0.02 & 0.012(2) & 0.6520(5)	&		38573(88) \\\hline
& 5 & 0.015 & 0.012(2) & 1.1514(8)	&		67833(149) \\\hline
& 5 & 0.01 & 0.013(2) &2.610(2)	&		152715(342) \\\hline
\end{tabular}
\end{table}

The simulation results support assumption that only the value of the last
jumps are important: the deviation of the hitting probability
$\Delta P_{\mathrm{av}}$ for $\delta_2=0.01$ is independent of the values of
both the parameters $R_c$ and $\delta_1$ (compare the last row for each
value of $R_c$). We can guess that it is reasonable to increase the value of
$\delta_1$ as much as possible, and the limit of $\delta_1$ from above is
$\delta_1\le R_c-R$. For example, we cannot use values of $\delta_1$ larger
than~1 for $R_c=11$ and $R=10$ or larger than~5 for $R_c=15$ and $R=10$ (see
Table~\ref{tab-simple}).

Results for $R_c=11$ and $\delta(r>R_c)=5$ are missing in the table. In this
case, a particle that is only 1~unit of length from the absorbing circle
$R=10$ makes a jump much larger than the distance to the circle, which
obviously causes huge errors in $P(\phi)$. In the simple algorithm, there is
a relation between the size of the region where small jumps are made and the
size of the large jump. Because we do not want the particle to jump from the
region $r>R_c$ to the absorbing circle $R$, we should ensure that
$R_c-R>\delta(r>R_c)>\delta(r\le R_c)$. The superior choice is
$R_c-R\gg\delta(r\le R_c)$, which results in a large number of steps in the
region close to the absorbing circle.

One can mention inspecting Table~\ref{tab-simple}, as well as the next two tables~\ref{tab-linear} and~\ref{tab-comparison}, that there are some minimum value of the deviation (approximately 0.013 like in the Table~\ref{tab-simple}). These can be explained
as the finite size effect of the finite number of bins, we use for the angle dependence of the hitting probability estimation.

\subsection{Linear protocol P3}

We next study the linear algorithm. Analysis of the simple algorithm shows
that we can increase jump length as much as possible while keeping some
region near the absorbing circle that is only accessible with small jumps.
The boundary case is $\delta(r)=r-R$, which means that the particle could
jump from any position in space to the absorbing sphere. We show results for
the linear algorithm in Table~\ref{tab-linear}.

\begin{table}[th!]
\centering
{\small
\caption{Performance and precision evaluation for linear protocol P3}
\label{tab-linear}
\begin{tabular}{|r|r|r|r|r|l|} \hline
$\delta_b$ & $\delta_h$ & $\Delta P_{\mathrm{av}}$ & $T$, min & $K$ \\
\hline\hline
1 & 0.3 & 0.025(2) & 0.01818(8) &787(2) \\\hline
1 & 0.2 & 0.022(2) & 0.01808(7)		&824(2) \\\hline
1 & 0.15 & 0.021(2) & 0.01927(8) 		&865(2) \\\hline
1 & 0.1 & 0.019(2) & 0.01932(8) 		&950(2)\\\hline
1 & 0.05 & 0.016(2) & 0.02163(8) 		&1134(2) \\\hline
1 & 0.03 & 0.012(2) & 0.02498(8) 		&1281(2) \\\hline
1 & 0.02 & 0.012(2) & 0.02677(7) 		&1419(2) \\\hline
1 & 0.01 & 0.010(2) & 0.02950(8) 		&1655(3) \\\hline
\hline
2 & 0.3 & 0.028(2) & 0.00494(6) 		&219.8(5) \\\hline
2 & 0.2 & 0.020(2) & 0.00498(6) 		&240.9(5) \\\hline
2 & 0.15 & 0.020(2) & 0.00543(7) 		&257.6(5) \\\hline
2 & 0.1 & 0.015(2)		& 0.00559(6) 	&283.5(5) \\\hline
2 & 0.05 & 0.016(2) & 0.00633(8) 		&336.9(6) \\\hline
2 & 0.03 & 0.016(2) & 0.00712(8) 		&378.9(5) \\\hline
2 & 0.02 & 0.013(2) & 0.00759(6) 		&414.7(6) \\\hline
2 & 0.01 & 0.013(2) & 0.00825(6)	&477.4(6) \\\hline
\hline
3 & 0.3 & 0.027(2) & 0.00238(9) 		&107.5(2) \\\hline
3 & 0.2 & 0.022(2) & 0.00252(9) 		&117.5(2) \\\hline
3 & 0.15 & 0.020(2) & 0.00265(9) 		&126.1(2) \\\hline
3 & 0.1 & 0.016(2)		& 0.00281(9) 	&139.1(3) \\\hline
3 & 0.05 & 0.015(2) & 0.00313(9) 		&163.8(3) \\\hline
3 & 0.03 & 0.012(2) & 0.00339(9) 		&183.4(3) \\\hline
3 & 0.02 & 0.012(2) & 0.00362(9) 		&199.5(3) \\\hline
3 & 0.01 & 0.012(2) & 0.00396(9)	&226.5(3) \\\hline
\hline
5 & 0.3 & 0.026(2) & 0.00101(7) 		&42.78(7) \\\hline
5 & 0.2 & 0.023(2) & 0.00107(7) 		&46.76(8) \\\hline
5 & 0.15 & 0.018(2) & 0.00111(7) 		&49.96(8) \\\hline
5 & 0.1 & 0.016(2)		& 0.00118(7) 	&54.61(9) \\\hline
5 & 0.05 & 0.015(2) & 0.00131(9) 		&63.25(9) \\\hline
5 & 0.03 & 0.014(2) & 0.00141(9) 		&69.9(1) \\\hline
5 & 0.02 & 0.013(2) & 0.00148(9) 		&75.4(1) \\\hline
5 & 0.01 & 0.015(2) & 0.00160(9)	&84.8(1) \\\hline
\hline
8 & 0.3 & 0.030(2) & 0.00044(7) 		&16.40(3) \\\hline
8 & 0.2 & 0.024(2) & 0.00046(7) 		&17.68(3) \\\hline
8 & 0.15 & 0.020(2) & 0.00048(6) 		&18.65(3) \\\hline
8 & 0.1 & 0.018(2) & 0.00051(6) 	&20.09(3) \\\hline
8 & 0.05 & 0.016(2) & 0.00055(6) 		&22.94(3)\\\hline
8 & 0.03 & 0.013(2) & 0.00059(7) 		&24.96(3) \\\hline
8 & 0.02 & 0.013(2) & 0.00062(7) 		&26.68(3) \\\hline
8 & 0.01 & 0.011(2) & 0.00065(7) 	&29.70(4) \\ \hline
\hline
16 & 0.3 & 0.298(2) & 0.00017(8) 		&5.51(2) \\\hline
16 & 0.2 & 0.299(2) & 0.00017(7) 		&5.50(2) \\\hline
16 & 0.15 & 0.295(2) & 0.00017(7) 		&5.53(2) \\\hline
16 & 0.1 & 0.299(2)		& 0.00018(7) 	&5.53(2) \\\hline
16 & 0.05 & 0.299(2) & 0.00018(7) 		&5.53(2) \\\hline
16 & 0.03 & 0.298(2) & 0.00018(8) 		&5.51(2) \\\hline
16 & 0.02 & 0.295(2) & 0.00018(7) 		&5.56(2) \\\hline
16 & 0.01 & 0.298(2) & 0.00018(8)	&5.54(2) \\\hline
\end{tabular}}%
\end{table}

The behavior of the linear algorithm is similar to the simple one.
Increasing the initial jump length $\delta(R_b)$ gives better performance,
and decreasing $\delta(R)$ gives better precision (and worse performance).
It is important that the jumps increase linearly and there is no upper
bound. Nevertheless, ratio $K_{\mathrm{return}}$ of trajectories that fly
away and are returned to $R_b$ is almost constant.

\subsection{Data analysis}

We compare the different algorithms in Table~\ref{tab-comparison}, where we
fixed the precision of the estimate of the hitting probability for a fair
comparison.

\begin{table}[!ht]
\centering
\caption{Comparison of algorithms with a precision equal to the precision of
the fixed jump length algorithm with $\delta=0.1$}
\label{tab-comparison}
\begin{tabular}{|l|r|r|r|r|r|r|} \hline
Algorithm & & & & $\Delta P_{\mathrm{av}}$ & $T$,min & Speedup \\
\hline
P1 & & $\delta=0.1$ & & 0.015(2) & 14.68(3) & \\
\hline \hline
P2 & $R_c$ & $\delta_1$ & $\delta_2$ & && \\ \hline
&11 & 1 & 0.03 & 0.015(2) & 0.1629(2) &88 \\ \hline
&15 & 1 & 0.01 & 0.014(2) &	 7.38(1) & 2 \\ \hline
&15 & 5 & 0.03 & 0.014(2) &0.2962(2) & 50 \\
\hline \hline
 P3& & $\delta_b$ & $\delta_h$ & & & \\
\hline
& & 1 & 0.05 & 0.016(2) & 0.02163(8) & 679 \\ \hline
& & 2 & 0.1 & 0.015(2)		& 0.00559(6) & 2626 \\\hline
& & 8 & 0.05 & 0.016(2) & 0.00055(6) & 26691 \\\hline
\end{tabular}
\end{table}

We choose the standard algorithm with $\delta=0.1$ as a reference and select
results for simple and linear algorithms that are close to it. The best
algorithm is the linear algorithm starting with $\delta=8$. This algorithm
is 20000 times faster than the algorithm with a fixed $\delta=0.1$ and 200
times faster than the algorithm with the fixed $\delta=1$.

The data in Tables~\ref{tab-simple} and \ref{tab-linear} can be analyzed to
obtain more information about the algorithm performance. The data in the
column Time in Table~\ref{tab-simple} can be fitted with a power law as a
function of $\delta_2$ with the exponent $y$,
\begin{equation}
T \propto \delta_2^{-y},
\label{fit-time}
\end{equation}
and the data in the column $K$ can be fitted with a power law with the
exponent $z$,
\begin{equation}
K \propto \delta_2^{-z}.
\label{fit-K}
\end{equation}

The fit of the data in Table~\ref{tab-simple} is presented in
Table~\ref{tab-time-2}. It is clear that the simulation time $T$ and the
number of steps increases as the second power of the inverse walk jump size
$\delta_2$.

\begin{table}[!ht]
\centering
\caption{Results of the fit to Eqs.~(\ref{fit-time}) and (\ref{fit-K}) for
the simple protocol P2.}
\label{tab-time-2}
\begin{tabular}{|l|l|l|l|} \hline
$R_c$ & $\delta_1$ & $y$ & $z$ \\
\hline
11 & 1& 2.52(2)	& 2.01(2)	\\ \hline
15 & 1& 2.047(1) & 2.05(2) \\ \hline
15 &5& 1.998(1) & 2.03(1) \\ \hline		
\end{tabular}
\end{table}

In the same manner, we can fit the data in Table~\ref{tab-linear} for the
linear algorithm (Protocol P3) by replacing $\delta_2$ with $\delta_b$:

\begin{eqnarray}
T &\propto& \delta_b^{-y} \\
\nonumber
K &\propto&\delta_b^{-z}.
\label{exponent3}
\end{eqnarray}

We show the results
of the fit in Table~\ref{tab-time-3}. Comparing Tables~\ref{tab-time-2}
and~\ref{tab-time-3}, we can see the drastic difference in the power-law
dependence for the simple protocol P2 and the linear protocol P3. The value
of the exponents $y\approx2$ and $z\approx2$ seems constant and rather large
in the case of the simple protocol P2. The values of the exponents for the
linear protocol P3 are quite smaller and seem saturated to the small value
$z\approx0.1$ (we do not have reliable values of the exponent $y$ in this
case).

\begin{table}[!ht]
\centering
\caption{Results of the fit to Eqs.~(\ref{fit-time}) and~(\ref{fit-K}) for
the linear protocol P3.}
\label{tab-time-3}
\begin{tabular}{|l|l|l|} \hline
$\delta_b$ & $y$ & $z$ \\
\hline
1& 0.34(1)	& 0.307(6)	\\ \hline
2& 0.17(4) & 0.174(5) \\ \hline
3& 0.15(1) & 0.132(5) \\ \hline
5& 0.057(20)	& 0.098(5)	\\ \hline
8& - & 0.097(6) \\ \hline		
\end{tabular}
\end{table}

\section{Discussions}
\label{sec-discussion}

We have numerically estimated the error in a random walk simulation. The
error is caused by the finite jump length that is not infinitesimally small
compared with the size of the absorbing circle. We calculated the error as a
function of the jump length $\delta$ and measured the angle-dependent
probability distribution. The deviation of the angle dependence could lead
to instabilities in a random cluster formation (e.g., in a DLA simulation).
We also tested the performance and precision of variable-jump-length
algorithms and showed that such algorithms can give a large performance
improvement, as can be seen comparing expressions ~(\ref{fit-time}), (\ref{fit-K}) and Table~\ref{tab-time-2} with expression~(\ref{exponent3}) and Table~\ref{tab-time-3}.

It should be noted that our results can be applied to the random walk only in two-dimensions. In larger dimensions, there is finite probability to escape to infinity while in two dimensions escaping probability is zero and random walk always return to the origin (despite the fact that the return time of the walk could be very large). In three dimensions, these leads to the interesting fact while looking for the probability that random walker will never be absorbed by the circle of radius $R$: the effective radius of the hitting sphere is changed linearly with the random walk jump size $\delta$. This effect was found by Ziff~\cite{DLA-3D} and in more details in the series of papers~\cite{MCZ06,ZMC07,ZMC09}. It is not clear how these results and the ones we describe in the present paper are connected.

\section{Acknowledgments}

This work has been initiated under the grant 14-21-00158 from Russian Science Foundation and finished within the ITP Landau research subject 0033-2019-0007.

\end{document}